\documentclass[%
reprint,
superscriptaddress,
aps,
pra,
notitlepage,
longbibliography,
]{revtex4-1}

\usepackage[utf8]{inputenc}

\usepackage{graphicx}
\usepackage{dcolumn}
\usepackage{bm}
\usepackage{mhchem}
\usepackage{xcolor}
\usepackage{enumerate}

\usepackage{natbib}
\usepackage[breaklinks,colorlinks=false,linkcolor=blue,urlcolor=blue,citecolor=red]{hyperref}

\usepackage{amsmath, esint, bbold, gensymb, amsfonts, amssymb, bm}
\allowdisplaybreaks

\usepackage{physics}
\usepackage{simpler-wick}
\usepackage{slashed} 
\usepackage[framemethod = tikz]{mdframed}
\usepackage[compat=1.1.0]{tikz-feynman}
\usepackage{tensor}
\usepackage{mhchem}
\usepackage{siunitx}

\newcommand{\D}{\mathcal{D}}

\renewcommand{\k}{\textbf{k}}

\renewcommand{\d}{\text{d}}
\DeclareMathOperator{\1}{\mathbb{1}}
\DeclareMathOperator{\SU}{SU}

\DeclareMathOperator{\algsu}{\mathfrak{su}}

\begin{document}

\title{Non-Trivial Fixed Points and Truncated SU(4) Kondo Models in a Quasi-Quartet Multipolar Quantum Impurity Problem}

\author{Daniel J. Schultz}
\affiliation{%
Department of Physics and Centre for Quantum Materials, University of Toronto, Toronto, Ontario, M5S 1A7, Canada
}%

\author{Adarsh S. Patri}
\affiliation{%
Department of Physics and Centre for Quantum Materials, University of Toronto, Toronto, Ontario, M5S 1A7, Canada
}%

\author{Yong Baek Kim}
\affiliation{%
Department of Physics and Centre for Quantum Materials, University of Toronto, Toronto, Ontario, M5S 1A7, Canada
}%

\date{\today}

\begin{abstract}
The multipolar Kondo problem, wherein the quantum impurity carries higher-rank multipolar moments, has seen recent theoretical and experimental interest due to proposals of novel non-Fermi liquid states and the availability of a variety of material platforms. The multipolar nature of local moments, in conjunction with constraining crystal field symmetries, leads to a vast array of possible interactions and resulting non-Fermi liquid ground states. Previous works on Kondo physics have typically focussed on impurities that have two degenerate internal states. In this work, inspired by recent experiments on the tetragonal material \ce{YbRu2Ge2}, which has been shown to exhibit a local moment with a quasi-fourfold degenerate ground state, we consider the Kondo effect for such a quasi-quartet multipolar impurity. In the tetragonal crystal field environment, the local moment supports dipolar, quadrupolar, and octupolar moments, which interact with conduction electrons in entangled spin and orbital states. Using renormalization group analysis, we uncover a number of emergent quantum ground states characterized by non-trivial fixed points. It is shown that these previously unidentified fixed points are described by truncated SU(4) Kondo models, where only some of the SU(4) generators (representing the impurity degrees of freedom) are coupled to conduction electrons. Such novel non-trivial fixed points are unique to the quasi-quartet multipolar impurity, reinforcing the idea that an unexplored rich diversity of phenomena may be produced by multipolar quantum impurity systems.
\end{abstract}

\maketitle

\section{\label{sec:introduction} Introduction}

Quantum impurity systems serve as important building blocks for understanding electron correlation effects in complex quantum materials and their theoretical models. For example, dynamical mean-field theory utilizes various quantum impurity systems as the mean-field frameworks to capture the local, but time-dependent, correlation effects \cite{Georges1996, Hu2020}. In mesoscopic physics, a quantum dot is often modeled as a quantum impurity which interacts with electrons in the leads of the device \cite{LeHur2015, Chang2009, Pustilnik2004}. The multi-channel Kondo effect, wherein multiple channels of conduction electrons interact with a local moment, leads to intriguing non-Fermi liquids \cite{Affleck1995, Gan1993, Gan1994a, Tsvelick1984a, Tsvelick1985a, Parcollet1997, Kimura2017, Bensimon2006, Cox1996, Affleck1991a}, offering the opportunity to study the interplay between non-Fermi liquids, broken symmetry states, and superconductivity in generalized lattice models of heavy fermion systems \cite{Doniach1977a, Cox1987a, Cox1988a, Cox1998a, Stewart1984, Fisk1985, Patri2021, Si2010, Si2013, MacLaughlin1984, Luke1993,Izawa2003,Ueda1987, Freeman1998}. In vast majority of examples, however, the quantum impurity is often modeled as a dipolar impurity characterized by a spin or a pseudospin $\frac{1}{2}$ quantum number. 

It has been recognized for some time that some local moments in $f$-electron systems carry higher-rank multipolar moments such as quadrupolar or octupolar moments \cite{Kusunose2008a, Kuramoto2009a, Thalmeier2008, Santini2009a}. In many of these systems, various non-Fermi liquid phenomena are observed \cite{Stewart2001, Yamane2018, Fu2020a}. While some aspects of multipolar local moments in a metallic host have been explored in the past, recent theoretical efforts have focussed on developing a deeper understanding of the interaction of conduction electrons (equipped with both spin and orbital degrees of freedom) scattering off higher-rank multipolar moments \cite{Patri2020d,Patri2020e,Schultz2021}. These unusual moments, arising due to crystalline electric field effects and strong spin-orbit coupling, transform in non-trivial ways under lattice group operations, and the abundance of possible moments, accompanied by a wealth of multi-orbital electrons, leads to a plethora of possible interactions and non-trivial ground states. Intriguingly, such systems have been shown to give rise to a range of exotic non-Fermi liquid behaviors. Common to all of these examples, however, is the fact that the local moment can access two internal states, for instance its spin up/down states in the case of the ordinary dipolar Kondo problem, or between (superpositions of) two higher total angular momentum states in certain non-Kramers doublet impurity systems. Recently, a further ingredient of diversity has been introduced with the discovery of a quartet of quasi-degenerate states for the localized $f$-electrons \cite{Jeevan2006, Rosenberg2019c}. The enhanced multiplicity of the localized states permits the development of many more multipolar moments than those appearing from (Kramers or non-Kramers) doublets. Constrained by the anisotropic crystalline electric field environment, these systems are a departure from the conventional quantum impurity problems, and the diverse landscape of higher-rank quantum impurities subject to a crystalline field remains largely unexplored. 

In this work, we consider a single quantum impurity with four quasi-degenerate crystal field levels in tetragonal systems. The quasi-quartet impurity hosts dipolar, quadrupolar, and octupolar moments, and we consider its symmetry-allowed interactions with conduction electrons belonging to the $E_u$ irreducible representation of the local $D_{4h}$ tetragonal symmetry. This physical setting is partly motivated by recent experimental studies on \ce{YbRu2Ge2} \cite{Rosenberg2019c, Pfau2019a, Ye2019a}. We classify the multipolar moments of the impurity according to irreducible representations of $D_{4h}$, and consider simplified models containing only the multipolar moments belonging to some closed sets of these irreps. Applying renormalization group techniques, we identify possible ground states, and analyze the low-energy thermodynamic and transport behaviors, such as specific heat and electrical resistivity. Specifically, when we consider multipolar moments in the $A_{1g}$ and $B_{1g}$ irreps, or the $A_{1g}$ and $B_{2g}$ irreps, we find a non-trivial fixed point characterized by a truncated $\SU(4)$ Kondo model, where only three generators of $\SU(4)$, representing the impurity degrees of freedom, are coupled to two channels of conduction electrons. When we consider multipolar moments from all of $A_{1g}$, $A_{2g}$, $B_{1g}$ and $B_{2g}$ irreps together, we obtain a different fixed point, characterized by a more complicated truncated $\SU(4)$ Kondo model, with six generators of $\SU(4)$ coupling to conduction electrons. These emergent fixed points are not known to be related to any known multi-channel Kondo models. In particular, the physical nature of the latter fixed point model intimates at the rearrangement of the multipolar moments into unusual emergent combinations, which results in different kinds of conduction electrons scattering with different local moment states. This work provides an illustrative example of the importance of the entangled spin/orbital nature of the conduction electrons due to the coupling to the multipolar impurity with multiple multipolar moments, and the novel ground states that result from such correlations.

The remainder of this paper is organized as follows. In Section \ref{sec:microscopic_models}, we explain the multipolar nature of the quasi-quartet ground state of the local moment and the conduction electrons which it scatters. In Section \ref{sec:kondo}, we introduce the symmetry-permitted multipolar Kondo Hamiltonians describing scattering of $E_u$-symmetry conduction electrons off of various choices of multipolar moments, namely moments belonging to the $A_{1g}$, $A_{2g}$, $B_{1g}$, and $B_{2g}$ irreducible representations of $D_{4h}$. In Section \ref{sec:rg}, we describe the low-energy behaviors of our models at renormalization group fixed points, calculate their scaling dimensions, and interpret the resulting fixed point models. In Section \ref{sec:discussions}, we explain the consequences of our work and discuss interesting future directions of exploration.

\section{Microscopic Models} \label{sec:microscopic_models}

\subsection{Quasi-Quartet Local Moment}
When a local moment is immersed in a crystalline electric field (CEF), the shapes of the electronic wave functions are restricted and can be classified according to the irreducible representations of the local point group. In the case of a \ce{Yb^{3+}} ion, a spin $J=7/2$ ground state is formed in the absence of a crystal field by Hund's rules \cite{Jeevan2006}. When this ion is subjected to a local tetragonal CEF, the 8 degenerate states split into four Kramers doublets. In this work, we consider the case where the two lowest-lying Kramers doublets feature an accidental quasi-degeneracy. Such a scenario is exemplified by the compound \ce{YbRu2Ge2} \cite{Ye2019a}. In this specific material, the splitting between the two lowest doublets is $T_0 = 10.2$ K, and the next excitation is separated by a further 360 K. At energy scales less than 360 K but not far below $T_0$, the four states can be taken to be equally thermally populated, and are all able to participate in Kondo scattering events; our analysis therefore applies in the range $10.2 \lesssim T < 360$ K. Indeed, measurements of the magnetic entropy reveal a plateau of the magnetic entropy at $k_B\log 4$ \cite{Jeevan2006}, indicating a quasi-quartet degeneracy of the $f$-electron levels and validating our theoretical setup. The quasi-quartet of states consists of linear combinations of $\ket{J,J_z}$ states from the $J=7/2$ multiplet:

\begin{equation} \label{eq:ground_states}
\ket{\Gamma_6^{(1)}}_\pm = a\ket{\frac{7}{2},\frac{\pm  7}{2}} + b\ket{\frac{7}{2},\frac{\mp 1}{2}},
\end{equation}

\begin{equation} \label{eq:first_excited_states}
\ket{\Gamma_7^{(1)}}_\pm = c\ket{\frac{7}{2}, \frac{\pm 5}{2}} + d\ket{\frac{7}{2},\frac{\mp 3}{2}},
\end{equation}

\noindent where, for \ce{YbRu2Ge2}, the experimentally determined values (from polarization resolved Raman scattering spectroscopy) are $a = -0.772$, $b = 0.636$, $c = 0.508$, and $d = 0.861$ \cite{Ye2019a}. This quasi-quartet supports 16 multipolar moments: 1 trivial identity, 3 dipoles, 5 quadrupoles, and 7 octopoles. The large number of states presents a very rich structure and allows for a plethora of possible couplings. The supported multipolar moments (in terms of Stevens operators) are presented in Table \ref{tb:stevens}.

\begin{table}
\centering
\begin{tabular}{c|c|c|c}
Projected & Abstract & Expression (action on $\ket{J,J_z}$) & Irrep \\ \hline 
${S}^0$ & $\mathcal{O}_0$ & $\1$ & $A_{1g}$ \\
${S}^1$ & ${J}_y$ & ${J}_y$ & $E_g$ \\
${S}^2$ & ${J}_z$ & ${J}_z$ & $A_{2g}$ \\
${S}^3$ & ${J}_x$ & ${J}_x$ & $E_g$ \\
${S}^4$ & ${\mathcal{Q}}_{2-}$ & $ -\sqrt{\frac{3}{5}} \frac{1}{24i}({J}_+^2 - {J}_-^2) $ & $B_{2g}$\\
${S}^5$ & ${\mathcal{Q}}_{1-}$ & $\frac{1}{2\sqrt{15}i}\overline{{J}_z({J}_+-{J}_-)}$ & $E_g$ \\
${S}^6$ & ${\mathcal{Q}}_0$ & $ -\frac{1}{6}(3{J}_z^2 - {\bm{J}}^2)$ & $A_{1g}$ \\
${S}^7$ & ${\mathcal{Q}}_{1+}$ & $ \frac{1}{2\sqrt{15}}\overline{{J}_z({J}_++{J}_-)} $ & $E_g$ \\
${S}^8$ & ${\mathcal{Q}}_{2+}$ & $ -\sqrt{\frac{3}{5}} \frac{1}{24}({J}_+^2 + {J}_-^2)$ & $B_{1g}$ \\
${S}^9$ & ${\mathcal{T}}_{3-}$ & $\frac{1}{60i} ({J}_+^3 - {J}_-^3)$ & $E_g$ \\
${S}^{10}$ & ${\mathcal{T}}_{2-}$ & $-\sqrt{\frac{3}{5}} \frac{1}{12i}\overline{{J}_z({J}_+^2 - {J}_-^2)} $ & $ B_{1g}$ \\
${S}^{11}$ & ${\mathcal{T}}_{1-}$ & $\frac{1}{20\sqrt{3}i}\overline{(5{J}_z^2 - {\bm{J}}^2 - 1/2)({J}_+-{J}_-)}$ & $E_g$ \\
${S}^{12}$ & ${\mathcal{T}}_0$ & $ \frac{1}{15}( 5{J}_z^3 - (3{\bm{J}}^2 - 1){J}_z)$ & $A_{2g}$ \\
${S}^{13}$ & ${\mathcal{T}}_{1+}$ & $ \frac{1}{20\sqrt{3}}\overline{(5{J}_z^2 - {\bm{J}}^2 - 1/2)({J}_++{J}_-)}$ & $E_g$ \\
${S}^{14}$ & ${\mathcal{T}}_{2+}$ & $-\sqrt{\frac{3}{5}}\frac{1}{12}\overline{{J}_z({J}_+^2+{J}_-^2)}$ & $B_{2g}$ \\
${S}^{15}$ & ${\mathcal{T}}_{3+}$ & $\frac{1}{60}({J}_+^3 + {J}_-^3)$ & $E_g$
\end{tabular}
\caption{The Stevens operators describing the identity ($\mathcal{O}$), dipolar (${J}$), quadrupolar (${\mathcal{Q}}$), and octupolar (${\mathcal{T}}$) moments are listed. We define the ``Abstract" notation to denote the microscopic operators which act on an arbitrary $\ket{J,J_z}$ state. The abstract operators have their explicit form given in the ``Expression" column. In the ``Projected" column, we present the notation for the Stevens operators after having been calculated in the quasi-quartet basis. The symmetry character according to the irreducible representations of $D_{4h}$ is listed in the ``Irrep" column. The overline notation denotes symmetrization $\overline{AB} := \frac{1}{2}(AB+BA)$. \label{tb:stevens}}
\end{table}

In a basis spanned by the four quasi-quartet states, these Stevens operators can be expressed in terms of the traceless $\SU(4)$ Gell-Mann generators and the identity operator. The explicit forms of the operators with respect to the particular basis given in Eqs. \eqref{eq:pseudo_basis_1} - \eqref{eq:pseudo_basis_4} are enumerated in Eqs. \eqref{eq:S0_matrix} - \eqref{eq:S15_matrix}. All multipole operators are linearly independent in this basis, despite some sharing the exact same symmetry. The fact that there are 16 multipolar moment $4\times 4$ matrices is to be contrasted with an ordinary $\SU(2)$ spin-3/2 moment, which, despite acting on a space spanned by four states, ($J_z = -\frac{3}{2}, -\frac{1}{2}, \frac{1}{2}, \frac{3}{2}$) possesses only three generators/moments. 

The multipolar operators in the quasi-quartet basis depend on $a,b,c,d$ in a highly convoluted form. To simplify the situation while capturing the essential low-energy behavior, we examine the case of $a=c=0$ and $b=d=1$ for the rest of this work. As can be seen in Appendix \ref{app:impurity}, the physics is preserved with this choice because the multipolar operators are all linearly independent (and nonzero) even for this simpler choice of wave function coefficients. 

\subsection{Conduction Electrons}
Conduction electrons in tetragonal systems can always be classified according to irreducible representations of $D_{4h}$. However, only when the electrons have a wave vector in the close vicinity of a high-symmetry point of the Brillouin zone, can we associate the basis states of irreducible representations with the degenerate energy levels of the conduction electrons. Recent ARPES studies indicated a small pocket in the Fermi surface about the $Z$ point of the first Brillouin zone in \ce{YbRu2Ge2} \cite{Pfau2019a}. This high-symmetry point leads to our consideration of two degenerate orbitals with $E_u$ symmetry; these two orbitals are labelled by $\{x,y\}$. The conduction electrons are able to transition between these two (symmetry-enforced) degenerate orbitals, as well as flip their spin, via interaction with the multipolar moments. 

\section{\label{sec:kondo} Multipolar Kondo Models}
We construct symmetry-allowed Hamiltonians that describe the scattering of conduction electrons with $x$ or $y$ symmetry off of the multipolar impurity. Because the impurity has 16 moments, we consider only certain subsets of multipolar operators to simplify the situation. Not every subset choice of pseudospin operators is allowable however; the set of operators must form a closed algebra dictated by symmetry. We use products of irreducible representations (see Table \ref{tb:d4h_irrep_prods}) to find candidate closed algebras by referring to Table \ref{tb:stevens}, where the pseudospin operators are classified by irreducible representation. In this work, we consider pseudospin operators belonging to the irreps $A_{1g}, A_{2g}, B_{1g}$, and $B_{2g}$. From these four irreps, we construct three intriguing models by considering the following closed multiplication tables:

\begin{enumerate}[(A)]
\item pseudospin operators in the $A_{1g}, B_{1g}$ irreps:
\begin{tabular}{c|cc}
$\otimes$ & $A_{1g}$ & $B_{1g}$ \\ \hline
$A_{1g}$ & $A_{1g}$ & $B_{1g}$ \\
$B_{1g}$ & $B_{1g}$ & $A_{1g}$
\end{tabular}
\item pseudospin operators in the $A_{1g}, B_{2g}$ irreps:
\begin{tabular}{c|cc}
$\otimes$ & $A_{1g}$ & $B_{2g}$\\ \hline
$A_{1g}$ & $A_{1g}$ & $B_{2g}$ \\
$B_{2g}$ & $B_{2g}$ & $A_{1g}$ 
\end{tabular}
\item pseudospin operators in the $A_{1g},A_{2g},B_{1g},B_{2g}$ irreps
\begin{tabular}{c|cccc}
$\otimes$ & $A_{1g}$ & $A_{2g}$ & $B_{1g}$ & $B_{2g}$ \\ \hline
$A_{1g}$ & $A_{1g}$ & $A_{2g}$ & $B_{1g}$ & $B_{2g}$ \\
$A_{2g}$ & $A_{2g}$ & $A_{1g}$ & $B_{2g}$ & $B_{1g}$ \\
$B_{1g}$ & $B_{1g}$ & $B_{2g}$ & $A_{1g}$ & $A_{2g}$ \\
$B_{2g}$ & $B_{2g}$ & $B_{1g}$ & $A_{2g}$ & $A_{1g}$
\end{tabular}
\end{enumerate}

\noindent Interaction of conduction electrons with the moments in these three tables will be the subject of the following three subsections.

\subsection{\label{sec:b1g_model} $A_{1g}\otimes B_{1g}$ Pseudospin Model}
The first model, as outlined in the introduction to this section, contains pseudospin operators in the $A_{1g}$ and $B_{1g}$ irreps. These operators are an identity (${S}^0$), two quadrupoles, (${S}^6,{S}^8$), and an octopole (${S}^{10}$). Indeed, the operators ${S}^6-{S}^0\in A_{1g}$, ${S}^8\in B_{1g}$, and ${S}^{10} \in B_{1g}$ form a canonically normalized $\algsu(2)$ algebra:

\begin{align} \label{eq:b1g_algebra}
[S^{10}, S^0-S^6] = i{S}^8,\\
[{S}^8, {S}^{10}] = i({S}^0-{S}^6), \\
[S^0-S^6, {S}^8] = i{S}^{10}.
\end{align}

\noindent These observations signify that, when conduction electrons scatter from multipolar moments, the localized $f$-electron must interchange between the particular internal states corresponding to the moments forming the closed subalgebra.  There are four Hamiltonians containing these pseudospins and respecting the full $D_{4h}$ symmetry and time-reversal symmetry:

\begin{align} 
H_{M} =& 2J_{M}c^\dagger_{0a\alpha} \sigma^z_{\alpha\beta} \tau^2_{ab}{S}^0 c_{0b\beta}, \label{eq:HM1}\\
H_{Q1} =& 2J_{Q1} c^\dagger_{0a\alpha}\sigma^z_{\alpha\beta} \tau^2_{ab}{S}^6 c_{0b\beta}, \label{eq:HQ1} \\
H_{Q2} =& 2J_{Q2} c^\dagger_{0a\alpha} \sigma^0_{\alpha\beta}\tau^3_{ab} {S}^8 c_{0b\beta}, \label{eq:HQ3} \\
H_{O1} =& 2J_{O1} c^\dagger_{0a\alpha}\sigma^z_{\alpha\beta}\tau^1_{ab} {S}^{10} c_{0b\beta} \label{eq:HO2}.
\end{align}

\noindent The Latin indices $a,b = \{x,y\}$ represent the conduction electrons' orbital degree of freedom, and the Greek indices $\alpha,\beta = \{\uparrow,\downarrow\}$ represent the spin degree of freedom. The $\sigma$ and $\tau$ matrices each form a canonically normalized $\algsu(2)$ Lie algebra (i.e. $[\sigma^i,\sigma^j] = i\epsilon^{ijk}\sigma^k$); we use different letters to avoid confusion between orbital $\tau$ and spin $\sigma$ spaces. Both the $a,b$ and $\alpha,\beta$ indices are summed over in these equations and other Hamiltonians in this work. The index 0 on the conduction electron operators indicates that these are operators at the origin (impurity site). These Hamiltonians are parameterized by four coupling constants, $J_{M}, J_{Q1}, J_{Q2},$ and $J_{O1}$. 

\subsection{\label{sec:b2g_model} $A_{1g}\otimes B_{2g}$ Pseudospin Model}
An analogous exercise can be conducted with other subsets of operators. In this case, we consider another closed algebra consisting of one identity (${S}^0$), two quadrupoles $({S}^4, {S}^6)$ and one octopole (${S}^{14}$). The operators ${S}^6-{S}^0$, ${S}^4$, and ${S}^{14}$ also form a closed $\algsu(2)$ subalgebra. The algebra is defined by the following commutation relations:

\begin{align}
[{S}^4, {S}^0-{S}^6] = i{S}^{14}, \\
[S^{14},{S}^4] = i({S}^0-{S}^6), \\
[{S}^0-{S}^6,{S}^{14}] = i{S}^4.
\end{align}

\noindent In similar fashion to the previous section, we construct a model wherein conduction electrons scatter off of moments in this closed subalgebra. The pseudospin operators are $\{{S}^0, {S}^6, {S}^4,{S}^{14}\}$, and this again leads to four Hamiltonians, with coupling constants $J_{M}, J_{Q1}, J_{Q3}$, and $J_{O2}$. The Hamiltonians for this model are Eqs. \eqref{eq:HM1} and \eqref{eq:HQ1}, along with two additional terms:

\begin{align} 
H_{Q3} =& 2J_{Q3} c^\dagger_{0a\alpha} \sigma^0_{\alpha\beta}\tau^1_{ab} {S}^4c_{0b\beta},\label{eq:HQ2} \\
H_{O2} =& 2J_{O2} c^\dagger_{0a\alpha}\sigma^z_{\alpha\beta}\tau^3_{ab}{S} ^{14} c_{0b\beta}. \label{eq:HO3} 
\end{align}

\noindent This set of 4 interaction Hamiltonians is isomorphic to those listed in Section \ref{sec:b1g_model}. The mapping for the isomorphism between the interactions can be done with the unitary transformation $c^\dagger_{x\uparrow} = a^\dagger_{1\uparrow}$, $c^\dagger_{y\uparrow} = a^\dagger_{2\uparrow}$, $c^\dagger_{x\downarrow} = a^\dagger_{2\downarrow}$, $c^\dagger_{y\downarrow} = -a^\dagger_{1\downarrow}$. The $a,a^\dagger$ operators are new fermionic operators which simply correspond to a new basis of conduction electrons.

\subsection{\label{sec:a2g_model} Combined Model}

The Stevens operators for the pseudospin form another closed algebra if one considers the operators in the $A_{2g}$ irrep, in addition to those in $A_{1g}$, $B_{1g}$ and $B_{2g}$ irreps. This appears organically because $B_{1g}\otimes B_{2g} = A_{2g}$, so combining the models of the previous two sections automatically generates these operators. Referring to Table \ref{tb:stevens}, we see that the two operators in the $A_{2g}$ representation are ${S}^2$ and ${S}^{12}$. The Hamiltonians that need to be introduced, in addition to the ones we already have (Eqs. \eqref{eq:HM1}-\eqref{eq:HO2}, and \eqref{eq:HQ3},\eqref{eq:HO3}), are given by 

\begin{align}
H_{D1} =& 2J_{D1}c^\dagger_{0a\alpha}\sigma^0_{\alpha\beta}\tau^2_{ab} {S}^2 c_{0b\beta}, \\
H_{D2} =& 2J_{D2}c^\dagger_{0a\alpha}\sigma^z_{\alpha\beta}\tau^0_{ab} {S}^2 c_{0b\beta}, \\
H_{O3} =& 2J_{O3} c^\dagger_{0a\alpha}\sigma^0_{\alpha\beta}\tau^2_{ab} {S}^{12} c_{0b\beta}, \\
H_{O4} =& 2J_{O4} c^\dagger_{0a\alpha}\sigma^z_{\alpha\beta}\tau^0_{ab} {S}^{12} c_{0b\beta}.
\end{align}

\noindent These models bring us to a grand total of 10 coupling constants. These new terms correspond to two dipole terms and two new octupolar terms. This combined model therefore has full participation of all (available) ranks of multipoles in the system.

\begin{figure*}[t]
\includegraphics[scale = 1.0]{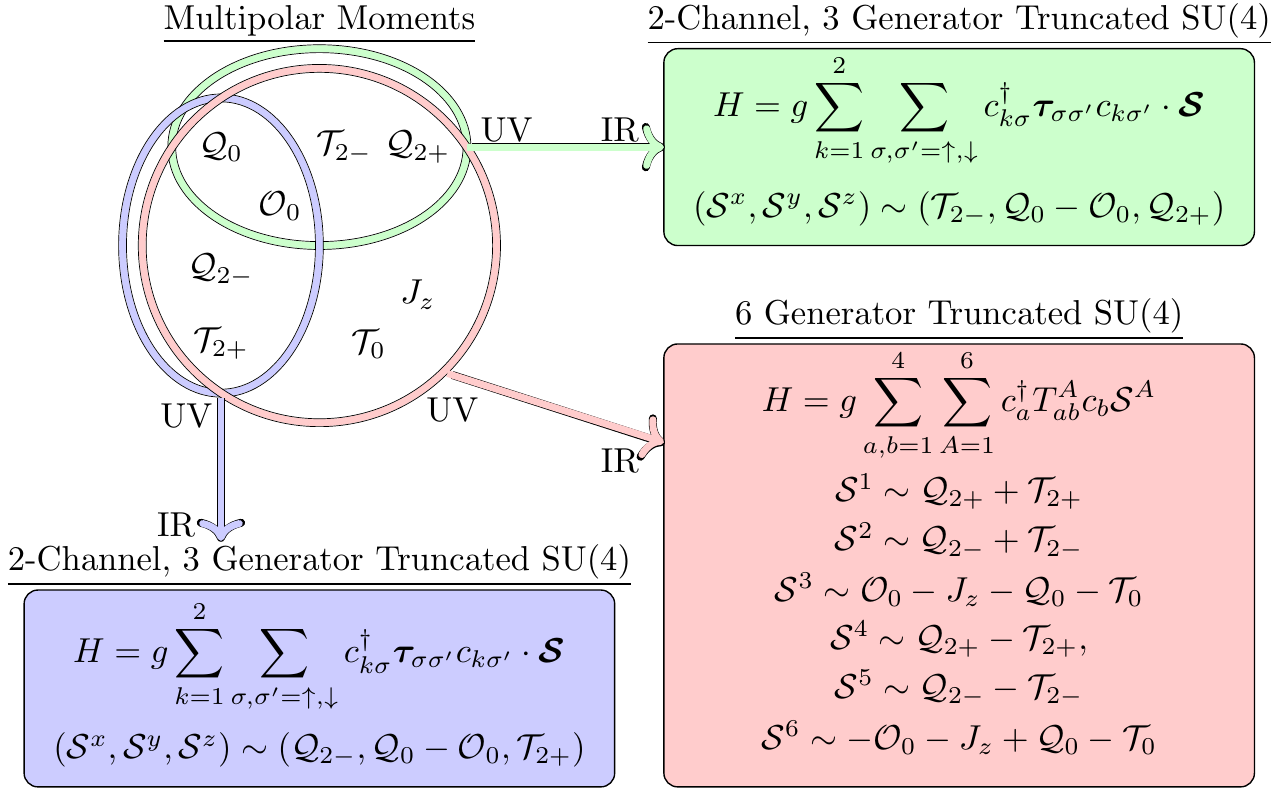}
\caption{Schematic denoting the renormalization group flow from the ultraviolet (UV) to infrared (IR) regime. In the UV limit, all the different moments have their own distinct couplings to conduction electrons, but they rearrange in the IR to emergent and simplified models. In the circles (top right), we consider different subsets of multipolar moments interacting with $x$ and $y$ conduction electrons. For the choice of $\{A_{1g},B_{2g}\}$ irrep moments, that is, $\{\mathcal{Q}_0,\mathcal{O}_0,\mathcal{Q}_{2-},\mathcal{T}_{2+}\}$ (blue), the IR theory becomes a 2-channel Kondo model with a truncated $\SU(4)$ spin (i.e. only 3 of the 15 generators of $\SU(4)$). For the $\{A_{1g}, B_{2g}\}$ irrep moments, $\{\mathcal{Q}_0,\mathcal{O}_0,\mathcal{Q}_{2+},\mathcal{T}_{2-}\}$ (red), the theory also becomes a 2-channel Kondo model with 3 generator truncated $\SU(4)$ moment in the IR. If the pseudospin is allowed more general transitions, we generate further symmetry-allowed interactions and the corresponding multipolar moments can lie in any of the irreps $\{A_{1g}, A_{2g}, B_{1g}, B_{2g}\}$ (red circle). In the IR limit, this yields a 6 generator truncated $\SU(4)$ Kondo model at the fixed point. There is a particularly substantial rearrangement of the moments, leadings us to define ``effective spins" $\mathcal{S}^A$ ($A=1,..6$) in terms of multipolar operators. In the ``2-channel" cases, $\mathcal{\tau}$ represents the standard $\algsu(2)$ Pauli matrices. In the 6 generator case, $T^A$, describing the conduction electrons, refers to 6 particular  $\SU(4)$ generators enumerated in the main text (Eq. \eqref{eq:fixed_point_hamil_6_gen}). This fixed point model can be interpreted as a single channel of $\SU(4)$ electrons interacting with a limited set of $\SU(4)$ moments.
} \label{fig:UV_to_IR}
\end{figure*}

\section{ \label{sec:rg} Nontrivial fixed points for Multipolar Kondo Models}

In this section, we will uncover the low energy behaviors of the models from Section \ref{sec:kondo} by calculating the scaling dimension $1+\Delta$ of the leading irrelevant operator for the different theories. We will find that the $A_{1g}\otimes B_{1g}$ and $A_{1g}\otimes B_{2g}$ critical models both have $\Delta_\text{perturb} = 1$, with two channels of conduction electrons scattering from only 3 generators of an $\SU(4)$ moment (as opposed to an $\SU(2)$ moment in the ordinary 2-channel Kondo effect). The third (combined) model possesses a new stable fixed point with 6 generators of the $\SU(4)$ moment coupling to the conduction electrons in a highly anistropic manner. These models are not known to be mappable to a $k$-channel Kondo model. We will discuss the stability and validity of these fixed points via a strong coupling analysis.

The constructed models require the renormalization group (RG) formalism in order to shine light onto the low-energy behaviors. In the standard Wilson RG, the interaction strengths depend on a high-energy cutoff $D$ corresponding to the bandwidth of the conduction electrons \cite{Anderson1970a,Wilson1975b}. As the cutoff is lowered, coupling constants flow, and eventually arrive at a fixed point. Stable fixed points correspond to the low-energy fate of the theory. The low-energy fixed points of the models are pictured schematically in Figure \ref{fig:UV_to_IR}.

The flow of the coupling constants is governed by $\beta$-functions, which we compute by expanding the vertex function to third order in perturbation theory. In contrast to previous works in Kondo problems, we need to carefully reconsider the wave function renormalization diagram, the details of which are outlined in Appendix \ref{app:feynman}. This additional complexity arises due to the anisotropic nature of the interaction, in addition to the $\algsu(4)$-type algebra obeyed by the pseudospin operators.

Physical information is contained within the $\beta$-functions via the slope $\Delta$ of the Jacobian at the fixed point, and the properties of resistivity and heat capacity scale as $\rho\sim T^\Delta$ and $C\sim T^{2\Delta}$ respectively \cite{Gan1993}. We note that the mentioned relationship of $\Delta$ to the physical quantities is not valid for the ordinary single-channel Kondo model, as there, the leading irrelevant operator is the bi-quadratic bosonic spin-current operator. This arises due to the impurity being completely absorbed into the bulk fermionic theory and rendering a mere change in the boundary condition \cite{Affleck1995}. Indeed, the striking feature of the single-channel Kondo model (which is distinct from $k$-channel Kondo models) is that the strong-coupling limit is non-degenerate, indicating the formation of the stable Kondo singlet. Below, we show that the critical Hamiltonians found here have degenerate ground states in the strong coupling limit, signifying the departure from this Fermi liquid picture. In the remainder of this section, we will discuss the fixed points and associated scaling behaviors for the models in question.

\subsection{\label{sec:b1g_b2g_fixed_points} $A_{1g}\otimes B_{1g}$ and $A_{1g}\otimes B_{2g}$ Models}
The $\beta$-functions for the coupling constants of the models in Sections \ref{sec:b1g_model} and \ref{sec:b2g_model} are given by:
\begin{align}
\frac{\d J_{M}}{\d\log D} =& - J_{O1}J_{Q2} - \frac{J_{Q1}}{2}\left(J_{O1}^2 + J_{Q2}^2\right),\label{eq:betaM1} \\
\frac{\d J_{Q1}}{\d\log D} =& J_{O1} J_{Q2} +  \frac{J_{Q1}}{2}\left(J_{O1}^{2} + J_{Q2}^{2}\right), \label{eq:betaQ1} \\
\frac{\d J_{Q2}}{\d\log D} =& J_{O1} J_{Q1} + \frac{J_{Q2}}{2}\left(J_{O1}^{2} + J_{Q1}^{2}\right), \label{eq:betaQ3} \\
\frac{\d J_{O1}}{\d\log D} =& J_{Q1} J_{Q2} + \frac{J_{O1}}{2}\left(J_{Q1}^{2} + J_{Q2}^{2}\right). \label{eq:betaO2}
\end{align}

\noindent We notice that the flow equation for $J_{M}$ is identical (up to a minus sign) to that for $J_{Q1}$. This indicates that $J_{M}$ can be chosen freely, because, upon solution of the other three $\beta$-functions, Eq. \eqref{eq:betaM1} will automatically be satisfied. Because the two models in Sections \ref{sec:b1g_model} and \ref{sec:b2g_model} are isomorphic to one another, we will analyze only the $A_{1g}\otimes B_{1g}$ model. Both sets of $\beta$-functions have four stable fixed points, each of which exhibits the same ground state with perturbative $\Delta_\text{perturb} = 1$. The explicit forms of the fixed points are presented in Appendix \ref{app:b1g_betas}. Both the $\beta$-functions and their solutions for the $A_{1g}\otimes B_{2g}$ model can be found by making the following substitutions in the results of the $A_{1g}\otimes B_{1g}$ model: $J_{O1}\to J_{O2}$, and $J_{Q2}\to J_{Q3}$.

In order to gain further insight into the models and nature of the fixed points, we approach the fixed point by defining a single coupling constant $g$, which, when flowing to $g\to 1$, yields the fixed point Hamiltonian. One of the RG fixed points is $(J_{M}, J_{Q1}, J_{Q2}, J_{O1}) = (1,-1,-1,-1)$, which can be obtained by setting $J_{M} = g, J_{Q1} = -g$, $J_{Q2} = -g$, $J_{O1} = -g$, and approaching $g\to 1$. The fixed point Hamiltonian takes the simplified form:

\begin{align}
H = 2gc^\dagger_{0a\alpha}\left[-\sigma^z_{\alpha\beta}\tau^1_{ab}S^{10} + \sigma^z_{\alpha\beta}\tau^2_{ab}(S^0-S^6) - \sigma^0_{\alpha\beta}\tau^3_{ab}S^8\right]c_{0b\beta}
\end{align}

\noindent In the strong coupling limit, $g\to\infty$, so the kinetic energy term can be ignored when compared to the Kondo interaction. The ground state of this strong coupling limit is in the two-particle sector and is fourfold degenerate. This indicates that additional scattering occurs at strong coupling, so the strong coupling limit is unstable, and the coupling constant $g$ flows back towards weak coupling \cite{Nozieres1980a}. It arrives at an intermediate fixed point, as the Gaussian fixed point is also unstable. These four degenerate ground states are listed in Appendix \ref{app:SC_b1g}. Further, when a perturbatively small kinetic term is introduced to allow coupling to neighbouring sites, the ground state remains fourfold degenerate, further signifying that the strong coupling limit is unstable. This demonstrates the validity and stability of the intermediate fixed point.

We can perform a change of basis on the conduction electrons to rewrite them in terms of a ``channel" degree of freedom and a ``spin" degree of freedom. The unitary transformation on the creation/annihilation operators is simply given by $c^\dagger_{x\uparrow} = \psi^\dagger_{1\downarrow}$, $c^\dagger_{y\uparrow} = -\psi^\dagger_{1\uparrow}$, $c^\dagger_{x\downarrow} = \psi^\dagger_{2\downarrow}$, $c^\dagger_{y\downarrow} = \psi^\dagger_{2\uparrow}$. The model is therefore rewritable as

\begin{align}
H = g\sum_{k=1}^2 \sum_{s',s = \uparrow,\downarrow}\psi^\dagger_{ks'}\bm{\sigma}_{s's}\psi_{ks}\cdot\textbf{S},
\end{align}

\noindent where $\textbf{S} = (S^{10}, S^0-S^6, S^8)$ and $\bm{\sigma} = (\sigma^x,\sigma^y,\sigma^z)$. The $\sigma^i$ matrices are canonically normalized $\algsu(2)$ Pauli matrices. Here, $k$ denotes the ``orbital" channel, and $s',s$ denote the ``spin" of the new operators. The distinction from the ordinary 2-channel $\SU(2)$ Kondo model is that the moment is replaced by an $\SU(4)$ moment. However, since a $\SU(4)$ moment has 15 generators, the number of available $\SU(4)$ moments is truncated; only 3 of the 15 generators are present. We emphasize that the three moments still satisfy the algebra in Eq. \eqref{eq:b1g_algebra}, which allows us to make an identification with a $\algsu(2)$ subalgebra of $\algsu(4)$. The fact that these are $\algsu(4)$ operators suggests that the exact solution for the $\SU(2)$ moment 2-channel model does not apply \cite{Affleck1995, Affleck1991a}. 

It is easier to see the $\SU(4)$ nature of the spin by writing the pseudospin operators in terms of Gell-Mann generators. Furthermore, the conduction electrons can also be written in terms of $\SU(4)$ generators, by using a different unitary basis change:

\begin{align}
c^\dagger_{x\uparrow} =& \frac{1}{\sqrt{2}}\left(-\chi^\dagger_{1} + \chi^\dagger_{2}\right) \label{eq:SOC_basis_1} \\
c^\dagger_{x\downarrow} =& \frac{1}{\sqrt{2}}\left(-\chi^\dagger_{3} + \chi^\dagger_{4}\right) \\
c^\dagger_{y\uparrow} =& \frac{i}{\sqrt{2}}\left(\chi^\dagger_{1} + \chi^\dagger_{2} \right) \\
c^\dagger_{y\downarrow} =& \frac{i}{\sqrt{2}}\left(\chi^\dagger_{3} + \chi^\dagger_{4}\right) \label{eq:SOC_basis_4}.
\end{align}

\noindent After this basis change, another structural layer of the model is revealed. If we write the pseudospin operators in the (ordered) basis 

\begin{equation} \label{eq:best_pseudospin_basis}
\left\{ \ket{\Gamma^{(1)}_7}_-, -\ket{\Gamma^{(1)}_6}_+, \ket{\Gamma^{(1)}_6}_-, -\ket{\Gamma^{(1)}_7}_+ \right\},
\end{equation}

\noindent then the spin operators (see Appendix \ref{app:impurity}) are expressed simply in terms of $\SU(4)$ Gell-Mann generators, defined explicitly in Appendix \ref{app:gell-mann}. Now, the generators of $\SU(4)$ which describe the conduction electrons clearly coincide with the generators describing the pseudospin,

\begin{align}
H =& g\sum_{a,b=1}^4\chi^\dagger_a\left[ (\Lambda^1_{ab} + \Lambda^{13}_{ab})(\lambda^1+\lambda^{13}) \right. \label{eq:fixed_point_hamil_2channel}\\
& \left. + (\Lambda^2_{ab}-\Lambda^{14}_{ab})(\lambda^2-\lambda^{14}) + (\Lambda^3_{ab}-\tilde{\Lambda}_{ab})(\lambda^3-\tilde{\lambda})\right]\chi_b \nonumber,
\end{align}

\noindent where $\tilde{\lambda} = \frac{1}{\sqrt{3}}(-\lambda^8 + \sqrt{2}\lambda^{15})$. We also define capitalized notation $\Lambda^A$ to represent the same $\SU(4)$ Gell-Mann generators for the conduction electrons; we have used a different letter to distinguish components of conduction electrons ($\Lambda$) from impurity operators ($\lambda$). The notation for the pseudospin operators can be read as $\lambda^A := \sum_{cd} f^\dagger_c \lambda^A_{cd} f_d$, where $f^\dagger$ and $f$ are Abrikosov psesudofermion operators, described in Appendix \ref{app:fedotov_popov}. Note also that this representation requires complex Fedotov-Popov chemical potentials to cancel out the unphysical Hilbert space sectors introduced by this parton construction, also described in Appendix \ref{app:fedotov_popov}. With the model written in this fashion, one can see that this fixed point model can alternatively be viewed as a highly limited $\SU(4)$ Kondo model. Indeed, the solution of such truncated $\SU(4)$ Kondo problems would depend precisely on which generators are chosen.

\subsection{\label{sec:combined_model_analysis} Combined Model}
The combined model of Section \ref{sec:a2g_model} had 10 coupling constants and the 10 resulting $\beta$-functions are given by:

\begin{widetext}
\begin{align}
\frac{\d J_{M}}{\d\log D} =& -\frac{\d J_{Q1}}{\d\log D} = -J_{O1}J_{Q2} - J_{O2}J_{Q3} - \frac{J_{Q1}}{2}(J_{Q3}^2 + J_{Q2}^2 + J_{O1}^2 + J_{O2}^2), \label{eq:a2g_beta1} \\
\frac{\d J_{D1}}{\d\log D} =& 2\frac{\d J_{O3}}{\d\log D} = 2J_{O1}J_{O2} + 2J_{Q3}J_{Q2} + (5J_{O3}-2J_{D1})(J_{Q3}^2+J_{Q2}^2+J_{O1}^2+J_{O2}^2), \label{eq:a2g_beta2} \\
\frac{\d J_{D2}}{\d\log D} =& 2\frac{\d J_{O4}}{\d\log D} = (5J_{O4}-2J_{D2})(J_{Q3}^2+J_{Q2}^2+J_{O1}^2+J_{O2}^2) , \label{eq:a2g_beta3} \\
\frac{\d J_{Q2}}{\d\log D} =& J_{O1}J_{Q1} + (5J_{O3} - 2J_{D1})J_{Q3} + \frac{J_{Q2}}{2}(J_{Q1}^2 + J_{Q3}^2 + J_{O1}^2 + (5J_{O3} - 2J_{D1})^2 + (5J_{O4} - 2J_{D2})^2 ), \label{eq:a2g_beta4} \\
\frac{\d J_{Q3}}{\d\log D} =& J_{O2}J_{Q1} + (5J_{O3}- 2J_{D1})J_{Q2}  + \frac{J_{Q3}}{2}(J_{Q1}^2 + J_{Q2}^2 + J_{O2}^2 + (5J_{O3} - 2J_{D1})^2 + (5J_{O4} - 2J_{D2})^2), \label{eq:a2g_beta5} \\
\frac{\d J_{O1}}{\d\log D} =& J_{Q1}J_{Q2} + (5J_{O3} - 2J_{D1})J_{O2} + \frac{J_{O1}}{2}(J_{O2}^2 + J_{Q1}^2 + J_{Q2}^2 + (5J_{O3} - 2J_{D1})^2 + (5J_{O4} - 2J_{D2})^2), \label{eq:a2g_beta6}\\
\frac{\d J_{O2}}{\d\log D} =& J_{Q1}J_{Q3} + (5J_{O3} - 2J_{D1})J_{O1} + \frac{J_{O2}}{2}(J_{O1}^2 + J_{Q1}^2 + J_{Q3}^2 + (5J_{O3} - 2J_{D1})^2 + (5J_{O4} - 2J_{D2})^2).\label{eq:a2g_beta7}
\end{align}
\end{widetext}

\noindent We note that there is a special grouping of coupling constants $J_{D1}$ and $J_{O3}$, and also $J_{D2}$ and $J_{O4}$. This can be understood because the Hamiltonians for these terms satisfy all the same symmetry properties; e.g. the $J_z$ dipole and $\mathcal{T}_{30}$ octopole are in the same irrep. We also note that (up to a constant multiple), the $\beta$-functions for $J_{D1}$ and $J_{O3}$ are the same, the $\beta$-functions for $J_{D2}$ and $J_{O4}$ are the same, and the $\beta$-functions for $J_{M}$ and $J_{Q1}$ are the same. Because three of the $\beta$-functions are redundant, we conclude that the space of solutions will be three-dimensional. At every point on this 3-dimensional solution manifold of fixed points, we find that with $\Delta_\text{perturb} = 2$. Points on the manifold are described by three free parameters $\zeta,\nu,\eta$. The origin of these free parameters and the explicit solution of the $\beta$-functions are shown in Appendix \ref{app:a2g_betas}. To study the fixed point Hamiltonian, we fix the ratios between the coupling constants by introducing a single parameter $g$, which, when taken to 1, lands on the perturbative fixed point. To gain insight into this fixed point Hamiltonian, we perform the same unitary change of basis on conduction electrons as in the previous section (Eqs. \eqref{eq:SOC_basis_1} - \eqref{eq:SOC_basis_4}). This decouples the conduction electrons into two sectors, which demonstrates that only scattering between $\{\chi_1,\chi_2\}$ and $\{\chi_3,\chi_4\}$ is allowed. In terms of the new $\chi$ operators and the free parameters describing the 3-dimensional manifold of solutions, the Hamiltonian becomes,

\begin{widetext}
\begin{align}
H =& g\begin{pmatrix} \chi^\dagger_1 & \chi^\dagger_2 \end{pmatrix}\left[\nu\tau^0\tilde{S}' + \tau^x(S^8 + S^{14}) + \tau^y(S^4 + S^{10}) + \tau^z(-S^6-\tilde{S}+\zeta \tilde{S}' + \eta S^0)\right]\begin{pmatrix}\chi_1 \\ \chi_2\end{pmatrix} \\
& +g\begin{pmatrix} \chi^\dagger_3 & \chi^\dagger_4 \end{pmatrix}\left[-\nu\tau^0\tilde{S}' + \tau^x(S^8 - S^{14}) + \tau^y(S^4 - S^{10}) + \tau^z(S^6-\tilde{S} + \zeta \tilde{S}' - \eta S^0)\right]\begin{pmatrix}\chi_3 \\ \chi_4 \end{pmatrix} \nonumber,
\end{align}
\end{widetext}

\noindent where the $\tau$ matrices are the canonically normalized $\algsu(2)$ Pauli matrices, and they are multiplied with the row and column vectors of $\chi$ conduction electron operators. For example, $\begin{pmatrix} \chi_1^\dagger & \chi_2^\dagger \end{pmatrix}\tau^0\tilde{S}'\begin{pmatrix}
	\chi_1 & \chi_2 \end{pmatrix}^T =\frac{1}{2}(\chi_1^\dagger \chi_1 + \chi_2^\dagger\chi_2)\tilde{S'}$. We have defined two special linear combinations of pseudospin operators for brevity,

\begin{equation}
\tilde{S} = -\frac{2}{29}S^2 + \frac{5}{29}S^{12},\qquad  \tilde{S}' = \frac{5}{29}S^2 + \frac{2}{29}S^{12}.
\end{equation}

\noindent The most convenient values of the fixed manifold parameters, $\zeta = -12,\nu = 0, \eta = 1$, maximally simplify the Hamiltonian. The simplification is made clear by writing the spin operators in the particular basis defined in the previous section Eq. \eqref{eq:best_pseudospin_basis}. For example, in this basis, $S^8+S^{14} = 2\lambda^1$; the other effective spin operators are written explicitly in terms of the Gell-Mann generators in Appendix \ref{sec:effective_moments}. In this simplified form, the Hamiltonian can be written as 

\begin{align}
H =& 2g\sum_{a,b=1}^4\chi_a^\dagger\left[\Lambda^1_{ab}\lambda^1+\Lambda^2_{ab}\lambda^2 + \Lambda^3_{ab}\lambda^3 \right. \label{eq:fixed_point_hamil_6_gen} \\
&\left. + \Lambda^{13}_{ab}\lambda^{13} + \Lambda^{14}_{ab}\lambda^{14} + \tilde{\Lambda}_{ab}\tilde{\lambda}\right]\chi_b. \nonumber
\end{align}

With the fixed point Hamiltonian in such a simple form, it is easier to analyze the strong-coupling limit $g\to\infty$. Diagonalizing the strong coupling Hamiltonian in all electron number sectors, we find a twofold degenerate ground state in the single-particle sector, a fourfold degenerate ground state in the two-particle sector, and a twofold degenerate ground state in the three-particle sector; the ground state energies of the one, two, and three particle sectors are all equal, whereas the ground states of the empty and fourfold occupied sectors are at a higher energy. The strong-coupling ground states are listed in Appendix \ref{app:SC_a2g}. Once again, these degeneracies indicate that the strong coupling limit is unstable, and the coupling constant must flow back towards an intermediate coupling. Additionally, perturbatively adding a kinetic term does not split the degeneracy of any of the single, double, or triple occupied sectors. This confirms the instability of the strong coupling limit and validity of the intermediate fixed point. 

Given the stability of the fixed point, we interpret the fixed point Hamiltonian in the following manner: depending on whether a $\chi_1,\chi_2$ electron or a $\chi_3,\chi_4$ electron scatters, the impurity electron will rearrange its occupation of the internal states (and the supported moments as well) into different forms, e.g. $S^8$ and $S^{14}$ become $S^8+S^{14}$ if a $\chi_1$ electron scatters, or $S^8-S^{14}$ if a $\chi_3$ scatters. These special combinations of multipolar moments constitute combined charge and magnetization densities, and therefore do not have definite character under time reversal. For example, electric quadrupole moments ($S^8$) are time-reversal even, and they are added here to magnetic octopoles ($S^{14}$), which are time-reversal odd.

Additionally, not every pseudospin can scatter with every conduction electron; this as can be seen by the representations of the effective spin operators in terms of Gell-Mann generators. Furthermore, the moment operators decouple into two closed $\algsu(2)$ subalgebras formed by $\{\lambda^1,\lambda^2,\lambda^3\}$ and $\{\lambda^{13},\lambda^{14},\tilde{\lambda}\}$. The upshot is that different flavours of conduction electrons couple to multipolar moments forming different sets of $\algsu(2)$ subalgebra.

Lastly, we want to contrast this fixed point model with the single-channel Kondo model. On the surface, $\Delta_\text{perturb} = 2$ artificially resembles the ($k=1$) single-channel Kondo model, because the $k$-channel perturbative scaling dimension for an $\SU(2)$ impurity is given in general by $\Delta_k = 2/k$. Despite this similarity, closer consideration revealed that this fixed point is not related to a single-channel Kondo model, or even a $k$-channel Kondo model. In particular, the strong coupling limit of our model is degenerate, whereas the strong-coupling limit of the single-channel model is non-degenerate (indicating the formation of the Kondo singlet). Thus the $\Delta_\text{perturb} = 2$ must have a different physical identity. Further evidence is encoded in the fact that, although the conduction interactions are interacting with the same impurity, they are interacting with different ``parts" of it; that is, different $f$-electron states (or multipolar moments) ``flip" depending on the flavour of the conduction electron that arrives into the impurity site.

\section{\label{sec:discussions} Discussions}
We have investigated a number of theoretical models for novel multipolar Kondo problems, where conduction electrons are interacting with a local moment in a quasi-quartet ground state arising from a tetragonal crystalline electric field. The quasi-quartet ground state leads to a plethora of multipolar moments carried by the local moment. Applying perturbative RG analysis, we uncovered the IR fixed points of these models and the scaling dimension of the leading irrelevant operators. Depending on the symmetry character of the multipolar moments we include, we found different non-trivial fixed points. When multipolar moments in the $\{A_{1g} , B_{1g} \}$ or $\{A_{1g} , B_{2g} \}$ irreps of $D_{4h}$ are considered, we found that the fixed point is a truncated $\SU(4)$ Kondo model, where only three generators of $\SU(4)$ describing the impurity states are coupled to two channels of conduction electrons. We also find that a leading irrelevant operator has the scaling dimension, $1+\Delta$, where the perturbative $\Delta_\text{perturb} = 1$. When we consider all of $A_{1g}$ , $A_{2g}$ , $B_{1g}$ , $B_{2g}$ irreps multipolar moments, we find a different non-trivial fixed point and a leading irrelevant operator with the scaling dimension, $1+\Delta$, with $\Delta_\text{perturb} = 2$. This fixed point is characterized by a distinct truncated $\SU(4)$ Kondo model, where six generators of $\SU(4)$ are now coupled to conduction electrons. At this fixed point, upon the scattering of conduction electrons, the 8 multipolar moments rearrange themselves into different effective moments depending on the flavour of the incoming electron. The strong coupling limits of the critical Hamiltonians at all of the above fixed points are unstable, which confirms the stability of these intermediate coupling fixed points. This result, combined with the fact that such critical Hamiltonians are not related to any known Kondo models, suggests that these fixed points represent novel ground states that were not identified before. Clearly the emergence of these fixed points results from the presence of multiple multipolar moments carried by the local impurity. The unfamiliar nature of the novel Kondo models found here presents an opportunity to explore unusual non-Fermi liquid behavior in multipolar impurity systems. 

We have analyzed this multipolar Kondo system in the context of a single impurity, especially with the compound \ce{YbRu2Ge2} in mind. This situation may be realizable experimentally by diluting the \ce{Yb^{3+}} ions, which constitute the $J = 7/2$ moments, with \ce{Lu^{3+}}. The full $4f$ shell of Lu does not support the higher rank moments. Indeed even upon dilution, the conduction electron band shape in the paramagnetic state remains largely intact, which further justifies the applicability of our model to diluted settings \cite{Pfau2019a, Nagashima2014a}. While \ce{YbRu2Ge2} possesses a lattice of multipolar impurities, experiments have observed an upturn in the resistivity at low temperatures, suggesting possible importance of the Kondo effect in understanding electronic properties \cite{Rosenberg2019c}.

The most natural direction of work going forward is to understand the precise nature of the nontrivial fixed points found in this work and the exact scaling dimension of the leading irrelevant operators. We are inspired by other works \cite{Affleck1993b,Affleck1995,Affleck1999, White1992, Holzner2009} to apply CFT or the density matrix renormalization group, to shed light on studying the models. Another direction would be the investigation of a model including multipolar moments in the $E_g$ representation as well. Such investigations could lead to a highly anisotropic version of $\SU(4)$ Kondo problem, where all generators are included. This would contrast previous works on the isotropic $\SU(N)$ Kondo model \cite{Bensimon2006, Kimura2017}, few of whose models have been inspired by a realistic compound. 

\section*{Acknowledgements}
This work was supported by NSERC of Canada and the Center for Quantum Materials at the University of Toronto. Y.B.K. is supported by the Killam Research Fellowship of the Canada Council for the Arts. 

\appendix

\begin{appendix}

\section{ \label{app:gell-mann} $\algsu(4)$ Gell-Mann Matrices}
We express the Stevens operators in terms of a quasi-quartet basis, described in Appendix \ref{app:impurity}. The easiest way to represent these matrices is as linear combinations of generators of $\SU(4)$, with an additional matrix corresponding to the identity; $\algsu(4)$ does not include the identity. These generators have been normalized to satisfy $\tr(\lambda^A\lambda^B) = \frac{1}{2}\delta_{AB}$, where $A,B=0,\dots,15$. For later use, we enumerate here the 16 Gell-Mann matrices, forming a basis for all $4\times 4$ Hermitian matrices:

\begin{alignat}{2}
\lambda^0 =& \frac{1}{2\sqrt{2}}\begin{pmatrix} 1 & 0 & 0 & 0 \\ 0 & 1 & 0 & 0 \\ 0 & 0 & 1 & 0 \\ 0 & 0 & 0 & 1 \end{pmatrix}, & \lambda^1 =& \frac{1}{2}\begin{pmatrix} 0 & 1 & 0 & 0 \\ 1 & 0 & 0 & 0 \\ 0 & 0 & 0 & 0 \\ 0 & 0 & 0 & 0 \end{pmatrix} \\
\lambda^2 =& \frac{1}{2}\begin{pmatrix} 0 & -i & 0 & 0 \\ i & 0 & 0 & 0 \\ 0 & 0 & 0 & 0 \\ 0 & 0 & 0 & 0 \end{pmatrix}, & \lambda^3 =& \frac{1}{2}\begin{pmatrix} 1 & 0 & 0 & 0 \\ 0 & -1 & 0 & 0 \\ 0 & 0 & 0 & 0 \\ 0 & 0 & 0 & 0 \end{pmatrix}\\
\lambda^4 =& \frac{1}{2}\begin{pmatrix} 0 & 0 & 1 & 0 \\ 0 & 0 & 0 & 0 \\ 1 & 0 & 0 & 0 \\ 0 & 0 & 0 & 0 \end{pmatrix}, & \lambda^5 =& \frac{1}{2}\begin{pmatrix} 0 & 0 & -i & 0 \\ 0 & 0 & 0 & 0 \\ i & 0 & 0 & 0 \\ 0 & 0 & 0 & 0 \end{pmatrix} \\
\lambda^6 =& \frac{1}{2}\begin{pmatrix} 0 & 0 & 0 & 0 \\ 0 & 0 & 1 & 0 \\ 0 & 1 & 0 & 0 \\ 0 & 0 & 0 & 0 \end{pmatrix}, & \lambda^7 =& \frac{1}{2}\begin{pmatrix} 0 & 0 & 0 & 0 \\ 0 & 0 & -i & 0 \\ 0 & i & 0 & 0 \\ 0 & 0 & 0 & 0 \end{pmatrix} \\
\lambda^8 =& \frac{1}{2\sqrt{3}}\begin{pmatrix} 1 & 0 & 0 & 0 \\ 0 & 1 & 0 & 0 \\ 0 & 0 & -2 & 0 \\ 0 & 0 & 0 & 0 \end{pmatrix}, & \lambda^9 =& \frac{1}{2}\begin{pmatrix} 0 & 0 & 0 & 1 \\ 0 & 0 & 0 & 0 \\ 0 & 0 & 0 & 0 \\ 1 & 0 & 0 & 0 \end{pmatrix} \\
\lambda^{10} =& \frac{1}{2}\begin{pmatrix} 0 & 0 & 0 & -i \\ 0 & 0 & 0 & 0 \\ 0 & 0 & 0 & 0 \\ i & 0 & 0 & 0 \end{pmatrix}, & \lambda^{11} =& \frac{1}{2}\begin{pmatrix} 0 & 0 & 0 & 0 \\ 0 & 0 & 0 & 1 \\ 0 & 0 & 0 & 0 \\ 0 & 1 & 0 & 0 \end{pmatrix} \\
\lambda^{12} =& \frac{1}{2}\begin{pmatrix} 0 & 0 & 0 & 0 \\ 0 & 0 & 0 & -i \\ 0 & 0 & 0 & 0 \\ 0 & i & 0 & 0 \end{pmatrix}, & \lambda^{13} =& \frac{1}{2}\begin{pmatrix} 0 & 0 & 0 & 0 \\ 0 & 0 & 0 & 0 \\ 0 & 0 & 0 & 1 \\ 0 & 0 & 1 & 0 \end{pmatrix} \\
\lambda^{14} =& \frac{1}{2}\begin{pmatrix} 0 & 0 & 0 & 0 \\ 0 & 0 & 0 & 0 \\ 0 & 0 & 0 & -i \\ 0 & 0 & i & 0 \end{pmatrix}, & \lambda^{15} =& \frac{1}{2\sqrt{6}}\begin{pmatrix} 1 & 0 & 0 & 0 \\ 0 & 1 & 0 & 0 \\ 0 & 0 & 1 & 0 \\ 0 & 0 & 0 & -3 \end{pmatrix}.
\end{alignat}

\noindent We have also used $\Lambda$ to denote the $\algsu(4)$ Gell-Mann matrices, and, as matrices, they are defined as $\Lambda^A = \lambda^A$. As mentioned in the main text (below Eq. \eqref{eq:fixed_point_hamil_2channel}), we have used an alternative symbol to distinguish matrices used for conduction electrons $(\Lambda)$ and impurity operators $(\lambda)$.

\section{\label{app:impurity} Quasi-Quartet Impurity}
When a $J=7/2$ moment is immersed in a tetragonal $D_{4h}$ crystal field, the eight degenerate states split. The experimentally determined ground states are explicitly listed in the main text; see Eqs. \eqref{eq:ground_states} and \eqref{eq:first_excited_states}. Out of these four states, we define a new orthonormal basis by reordering and adding phases. This is the same basis as defined in Eq. \eqref{eq:best_pseudospin_basis}, but with the $\ket{\varphi}$ notation for compactness:

\begin{align} 
\ket{\varphi_1} =& \ket{\Gamma^{(1)}_7}_-\label{eq:pseudo_basis_1},\\
\ket{\varphi_2} =&  -\ket{\Gamma^{(1)}_6}_+ \label{eq:pseudo_basis_2}, \\
\ket{\varphi_3} =& \ket{\Gamma^{(1)}_6}_-\label{eq:pseudo_basis_3}, \\
\ket{\varphi_4} =& -\ket{\Gamma^{(1)}_7}_+ \label{eq:pseudo_basis_4}.
\end{align}

\noindent In this basis, we can project the Stevens operators, which normally act on an 8-element $J=7/2$ basis $\ket{J,m}$, to act on this 4-element basis Eqs \eqref{eq:pseudo_basis_1} - \eqref{eq:pseudo_basis_4}. In terms of the normalized $\algsu(4)$ generators defined by Appendix \ref{app:gell-mann}, these project Stevens operators take the form given in Eqs. \eqref{eq:S0_matrix} - \eqref{eq:S15_matrix}:

\begin{widetext}
\begin{align}
S^0 =& 4 \sqrt{2}\lambda^0 \label{eq:S0_matrix}\\
S^1 =&  (\sqrt{7} a c + \sqrt{15} b d)(\lambda^5+\lambda^{12}) + 4 b^{2}\lambda^7 + 4 \sqrt{3} c d\lambda^{10} \\
S^2 =& (-7a^2+b^2)\left(\frac{\lambda^3}{2} - \frac{\sqrt{3}}{2}\lambda^8\right) + (-5c^2+3d^2)\left(\frac{\lambda^3}{2} + \frac{\sqrt{3}}{6}\lambda^8 + \frac{\sqrt{6}}{3}\lambda^{15}\right) \\
S^3 =& (\sqrt{7} a c + \sqrt{15} b d)(\lambda^4+\lambda^{11})  - 4 b^{2}\lambda^6 - 4 \sqrt{3} c d\lambda^9 \\
S^4 =& \left(- \frac{\sqrt{35} a d}{10} - \frac{\sqrt{3} b c}{2} + b d\right)(\lambda^2+\lambda^{14}) \\
S^5 =& \left(- \frac{\sqrt{105} a c}{5} + b d\right)(\lambda^5-\lambda^{12}) \\
S^6 =& (-7a^2+5b^2)\left(\frac{\sqrt{2}}{2}\lambda^0 - \frac{\lambda^3}{2} - \frac{\sqrt{3}}{6}\lambda^8 + \frac{\sqrt{6}}{6} \lambda^{15}\right) + (-c^2+3d^2)\left(\frac{\sqrt{2}}{2}\lambda^0 + \frac{\lambda^3}{2} + \frac{\sqrt{3}}{6}\lambda^8 - \frac{\sqrt{6}}{6}\lambda^{15}\right) \\
S^7 =& \left( - \frac{\sqrt{105} a c}{5} + b d\right)(\lambda^4-\lambda^{11}) \\
S^8 =& \left( \frac{\sqrt{35} a d}{10} + \frac{\sqrt{3} b c}{2} + b d\right)(\lambda^1+\lambda^{13}) \\
S^9 =& -\frac{4 \sqrt{5} b c}{5}(\lambda^5+\lambda^{12}) - \frac{2 \sqrt{35} a b}{5}\lambda^7  - 2 d^{2}\lambda^{10} \\
S^{10} =& \left( - \frac{\sqrt{35} a d}{2} + \frac{3 \sqrt{3} b c}{2} + b d\right)(\lambda^2-\lambda^{14}) \\
S^{11} =& (\sqrt{21} a c - \sqrt{5} b d)(\lambda^5+\lambda^{12})  - 2 \sqrt{3} b^{2}\lambda^7 +  2 c d\lambda^{10} \\
S^{12} =& (7a^2+3b^2)\left(-\frac{\lambda^3}{2} + \frac{\sqrt{3}}{2}\lambda^8\right) + (5c^2-7d^2)\left(\frac{\lambda^3}{2} + \frac{\sqrt{3}}{6}\lambda^8 + \frac{\sqrt{6}}{3}\lambda^{15}\right) \\
S^{13} =& (\sqrt{21} a c - \sqrt{5} b d)(\lambda^4+\lambda^{11}) + 2 \sqrt{3} b^{2}\lambda^6  - 2 c d\lambda^9 \\
S^{14} =& \left(\frac{\sqrt{35} a d}{2} - \frac{3 \sqrt{3} b c}{2} + b d\right)(\lambda^1-\lambda^{13}) \\
S^{15} =& \frac{4 \sqrt{5} b c}{5}(\lambda^4+\lambda^{11}) - \frac{2 \sqrt{35} a b}{5}\lambda^6 - 2 d^{2}\lambda^9 \label{eq:S15_matrix}
\end{align}
\end{widetext}

\section{\label{app:symmetries} Symmetry Analysis of $D_{4h}$ }
In order to construct Hamiltonians respecting the local tetragonal symmetry, we need to find out how each constituent degree of freedom transforms under the generators of this group. The three generators we pick are a $\mathcal{C}_4$, a rotation by $\pi/2$ about the $z$ axis, $\mathcal{C}_2$, a rotation by $\pi$ about the $x$-axis, and $\mathcal{I}$, spatial inversion. We also include time reversal symmetry. The results are in Table \ref{tb:d4h_transformations}.

\begin{table}
\centering
\begin{tabular}{c|c|c|c|c}
Object & $\mathcal{C}_4$ & $\mathcal{C}_2'$ & $\mathcal{I}$ & $\mathcal{T}$ \\ \hline
$x$ & $y$ & $x$ & $-x$ & $x$ \\
$y$ & $-x$ & $-y$ & $-y$ & $y$ \\
$z$ & $z$ & $-z$ & $-z$ & $z$ \\
$\sigma^0$ & $\sigma^0$ & $\sigma^0$ & $\sigma^0$ & $\sigma^0$ \\
$\sigma^x$ & $\sigma^y$ & $\sigma^x$ & $\sigma^x$ & $-\sigma^x$ \\
$\sigma^y$ & $-\sigma^x$ & $-\sigma^y$ & $\sigma^y$ & $-\sigma^y$ \\
$\sigma^z$ & $\sigma^z$ & $-\sigma^z$ & $\sigma^z$ & $-\sigma^z$ \\
${S}^0$ & ${S}^0$ & ${S}^0$ & ${S}^0$ & ${S}^0$ \\
${S}^1$ & $-{S}^3$ & $-{S}^1$ & ${S}^1$ & $-{S}^1$ \\
${S}^2$ & ${S}^2$ & $-{S}^2$ & ${S}^2$ & $-{S}^2$ \\
${S}^3$ & ${S}^1$ & ${S}^3$ & ${S}^3$ & $-{S}^3$ \\
${S}^4$ & $-{S}^4$ & $-{S}^4$ & ${S}^4$ & ${S}^4$ \\
${S}^5$ & $-{S}^7$ & ${S}^5$ & ${S}^5$ & ${S}^5$ \\
${S}^6$ & ${S}^6$ & ${S}^6$ & ${S}^6$ & ${S}^6$ \\
${S}^7$ & ${S}^5$ & $-{S}^7$ & ${S}^7$ & ${S}^7$ \\
${S}^8$ & $-{S}^8$ & ${S}^8$ & ${S}^8$ & ${S}^8$ \\
${S}^9$ & ${S}^{15}$ & $-{S}^9$ & ${S}^9$ & $-{S}^9$ \\
${S}^{10}$ & $-{S}^{10}$ & ${S}^{10}$ & ${S}^{10}$ & $-{S}^{10}$ \\
${S}^{11}$ & $-{S}^{13}$ & $-{S}^{11}$ & ${S}^{11}$ & $-{S}^{11}$ \\
${S}^{12}$ & ${S}^{12}$ & $-{S}^{12}$ & ${S}^{12}$ & $-{S}^{12}$ \\
${S}^{13}$ & ${S}^{11}$ & ${S}^{13}$  & ${S}^{13}$ & $-{S}^{13}$ \\
${S}^{14}$ & $-{S}^{14}$ & $-{S}^{14}$ & ${S}^{14}$ & $-{S}^{14}$ \\
${S}^{15}$ & $-{S}^9$ & ${S}^{15}$ & ${S}^{15}$ & $-{S}^{15}$ 
\end{tabular}
\caption{Transformations under $D_{4h}$ and time reversal $\mathcal{T}$ of all relevant objects in the Hamiltonian. Here, $\sigma^i$ are the Pauli spin matrices with identity, normalized with $[\sigma^i,\sigma^j] = i\epsilon^{ijk}\sigma^k$, and the ${S}^i$ are the pseudospin operators.} \label{tb:d4h_transformations}
\end{table}

\subsection{Irreducible Representation Products}
When constructing models, it is useful to understand products of the irreducible representations. For the reader's convenience, we list the table of products of irreducible representations for the relevant irreps of $D_{4h}$ in Table \ref{tb:d4h_irrep_prods}. 

\begin{table}[ht]
\centering
\begin{tabular}{c|ccccc}
$\otimes$ & $A_{1g}$ & $A_{2g}$ & $B_{1g}$ & $B_{2g}$ & $E_g$ \\ \hline
$A_{1g}$ & $A_{1g}$ & $A_{2g}$ & $B_{1g}$ & $B_{2g}$ & $E_g$ \\
$A_{2g}$ & $A_{2g}$ & $A_{1g}$ & $B_{2g}$ & $B_{1g}$ & $E_g$ \\
$B_{1g}$ & $B_{1g}$ & $B_{2g}$ & $A_{1g}$ & $A_{2g}$ & $E_g$ \\
$B_{2g}$ & $B_{2g}$ & $B_{1g}$ & $A_{2g}$ & $A_{1g}$ & $E_g$ \\
$E_g$ & $E_g$ & $E_g$ & $E_g$ & $E_g$ & $A_{1g}\oplus[A_{2g}]\oplus B_{1g}\oplus B_{2g}$
\end{tabular}
\caption{Products of Irreducible Representations of $D_{4h}$}\label{tb:d4h_irrep_prods}
\end{table}

\section{\label{app:fedotov_popov} Generalized Fedotov-Popov Trick}
When doing perturbation theory, one often needs to expand $n$-point correlation functions in terms of 2-point ones. Wick's theorem presents a solution to this, but only for operators obeying canonical commutation or anticommutation relations. Spin operators obey a Lie algebra, so Wick's theorem does not apply. It is possible however to represent spin operators in terms of Abrikosov pseudofermion operators, obeying canonical anticommutation relations, which circumvent this difficulty. However, this representation artificially expands the Hilbert space on which these spin operators act, the notion of occupation number now being well defined. In order to restrict to the original Hilbert space, we require that the contribution to the partition function of sectors of occupation other than 1 all cancel out among themselves. This requires extension of the Fedotov-Popov trick for $\SU(2)$ spins, which normally cancels out the empty and doubly occupied pseudofermion sectors, to an $\SU(N)$ spin \cite{Veits1994}. An $\SU(N)$ spin, in the fundamental representation, can have up to $N$-states occupied when represented by pseudofermions. Furthermore, an $\SU(N)$ spin can be represented on a vector space of dimension other than $N$, so to maintain full generality we let $d$ be the dimension of the representation space of the spin. This allows $d+1$ possible occupations, from empty, 0, up to full, $d$. For example, $d = 2s+1$ for a spin $s$ $\algsu(2)$ spin, or, $d=N$ for an $\algsu(N)$ spin in the fundamental representation. To cancel out $d$ of these sectors and leave the singly occupied sector untouched in the partition function, we introduce $d$ complex chemical potentials,

\begin{equation}
	\mu_\ell = \frac{i\pi}{\beta}\frac{2\ell-1}{d}, \qquad \ell = 1,\dots,d. 
\end{equation}

\noindent This choice of chemical potentials means the equivalence of the ordinary spin partition function with the spin partition function written as a path integral over the Abrikosov pseudofermion fields. To do this, we consider the effective action 

\begin{equation}
S_0^f = \int_0^\beta \left[ \sum_{\gamma=1}^d\bar{f}_\gamma(\tau)\partial_\tau f_\gamma(\tau) + H_0^f(\bar{f}(\tau),f(\tau))\right] \d\tau.
\end{equation}

\noindent The free pseudofermion Hamiltonian has solely the chemical potential:

\begin{equation}
H_0^f(\mu) = -\mu\sum_{\gamma=1}^d \bar{f}_\gamma(\tau)f_\gamma(\tau).
\end{equation}

\noindent We Fourier transform the effective action to write it in terms of the fermionic Matsubara frequencies:

\begin{align}
f_\gamma(\tau) =& \frac{1}{\sqrt{\beta}}\sum_{i\omega_n} f_\gamma(i\omega_n)e^{-i\omega_n\tau}\\	
\bar{f}_\gamma(\tau) =& \frac{1}{\sqrt{\beta}}\sum_{i\omega_n} \bar{f}_\gamma(i\omega_n)e^{i\omega_n\tau}.
\end{align}

\noindent The action then becomes 

\begin{align}
S_0^f =& \int_0^\beta \sum_{\gamma=1}^d \left[ \bar{f}_\gamma(\tau)\partial_\tau f_\gamma(\tau) - \mu \bar{f}_\gamma(\tau)f_\gamma(\tau) \right] \d\tau \\
=& -\sum_{\gamma=1}^d\sum_{i\omega_n} \bar{f}_\gamma(i\omega_n)f_\gamma(i\omega_n)(i\omega_n + \mu).
\end{align}

\noindent At this point, we have not used the different complex chemical potentials in the calculation yet. In fact, the different chemical potentials do not enter in to the Hamiltonian at all, but rather when correlation functions are calculated. This allows us to calculate the free partition function $Z^f_0$ for the pseudofermions at a particular chemical potential. We do this simply by calculating 

\begin{align}
Z_0^f(\mu) =& \int e^{-S_0^f} \D\bar{f} \D f \\
=& (1+e^{\beta\mu})^d \label{eq:fock_partition_function} .
\end{align}

\noindent This is the typical result for a fermionic partition function for $d$ different occupiable states. However, this partition function came from a trace over all different occupation number sectors for the pseudofermions. To calculate the true partition function $Z_\text{spin}$, we restrict to the singly occupied sector:

\begin{align}
Z_\text{spin} =& \tr_{n=1} e^{-\beta H} \\
=& \sum_{n=0}^d \delta_{n,1}\tr_n e^{-\beta H}
\end{align}

\noindent where $n$ is the occupation number, and $\tr_n$ is the trace over all states with $n$ particles, and $d$ is the maximum occupation number. By writing the Kronecker delta in terms of its Fourier transform 

\begin{equation}
\delta_{n,1} = \frac{1}{d}\sum_{\ell=1}^d e^{\beta \mu_\ell	(n-1)},
\end{equation}

\noindent we can express the true partition function as 

\begin{align}
Z_\text{spin} =& \sum_{n=0}^d \frac{1}{d}\sum_{\ell=1}^d e^{\beta\mu_\ell(n-1)}\tr_n e^{-\beta H} \\
=& \frac{1}{d}\sum_{\ell=1}^d e^{-\beta \mu_\ell}\sum_{n=0}^d\tr_n\left(e^{-\beta(H-\mu_\ell n)}\right) \\
=& \frac{1}{d}\sum_{\ell=1}^d e^{-\beta\mu_\ell}Z_0^f(\mu_\ell)
\end{align}

\noindent Recognizing this alternative way of writing $Z^f_0(\mu_\ell)$, we can plug in its form calculated from the path integral version in Eq. \eqref{eq:fock_partition_function}. Thus the true partition function is then defined in terms of the Fock space partition function at different chemical potentials:

\begin{align}
Z_\text{spin} =& \frac{1}{d}\sum_{\ell=1}^d e^{-\mu_\ell\beta} (1+e^{\beta\mu_\ell})^d\\
=& d.
\end{align}

\section{Abrikosov Pseudofermion Green Function}
Correlation functions of Abrikosov pseudofermions need to be calculated with respect to the true partition function, described in the previous section. However, even introducing the Abrikosov pseudofermions requires us to use the definition of correlation functions on the Fock space. We relate the two quantities in an analogous way to how the partition functions are related:

\begin{align}
\mathcal{G}^f_0(i\omega_n) =& -\left<f_\gamma(i\omega_n)\bar{f}_\gamma(i\omega_n)\right>_0 \\
:=& \frac{1}{Z_\text{spin} d}\sum_{\ell=1}^d e^{-\mu_\ell\beta} \left(-Z^f_0(\mu_\ell)\left<f_\gamma(i\omega_n)\bar{f}_\gamma(i\omega_n)\right>_{\mu_\ell,0}\right)\\
=& \frac{1}{d^2}\sum_{\ell=1}^d e^{-\mu_\ell\beta} (1+e^{\beta\mu_\ell})^d \frac{1}{i\omega_n + \mu_\ell}. \label{eq:pseudofermion_propagator}
\end{align}

\noindent where the Fock space correlation function is defined in the usual way:

\begin{equation}
\left<A\right>_{\mu,0} := \frac{1}{Z^f_0(\mu)}\int A e^{-S^f_0} \D\bar{f}\D f.
\end{equation}

\noindent In this paper, $d=4$ because the quasi-quartet has a 4-element basis; in other words we are working with an $\algsu(4)$ spin.

\section{\label{app:feynman} Diagrammatic Perturbation Theory}
In order to calculate the $\beta$-functions for the theory, we calculate the vertex function to 2-loop order. Only the irreducible diagrams renormalize the interaction couplings in the renormalization group flow. In contrast to $\SU(2)$ Kondo problems, the generators $\lambda^A$ of $\SU(N)$ do not satisfy $(\lambda^A)^2$ is a Casimir operator. This means that the $1/3!$ from the Dyson expansion is partially cancelled by a factor of 3 from 3 choices of contracting the first conduction electron operator with an external point, but contracting the next operator yields a different result for the wave function renormalization diagrams. This introduces an additional wave function renormalization diagram for anisotropic $\SU(N)$ problems, and therefore the $1/2$ for wave function renormalization diagrams does not cancel out. The symmetry remains for the vertex correction diagram and cancels the remaining $1/2$. We have enumerated all relevant diagrams below in Figure \ref{fig:feynman}. In the figure, the solid lines are the free fermion propagator

\begin{equation}
\mathcal{G}^\text{c}_0(\k,i\omega_n) = \frac{1}{i\omega_n - \xi_\k} \label{eq:cond_prop} 
\end{equation}

\noindent and the dashed lines are the pseudofermion propagator from Eq. \eqref{eq:pseudofermion_propagator}.

\begin{figure}[ht]%
\centering
\includegraphics[scale = 1]{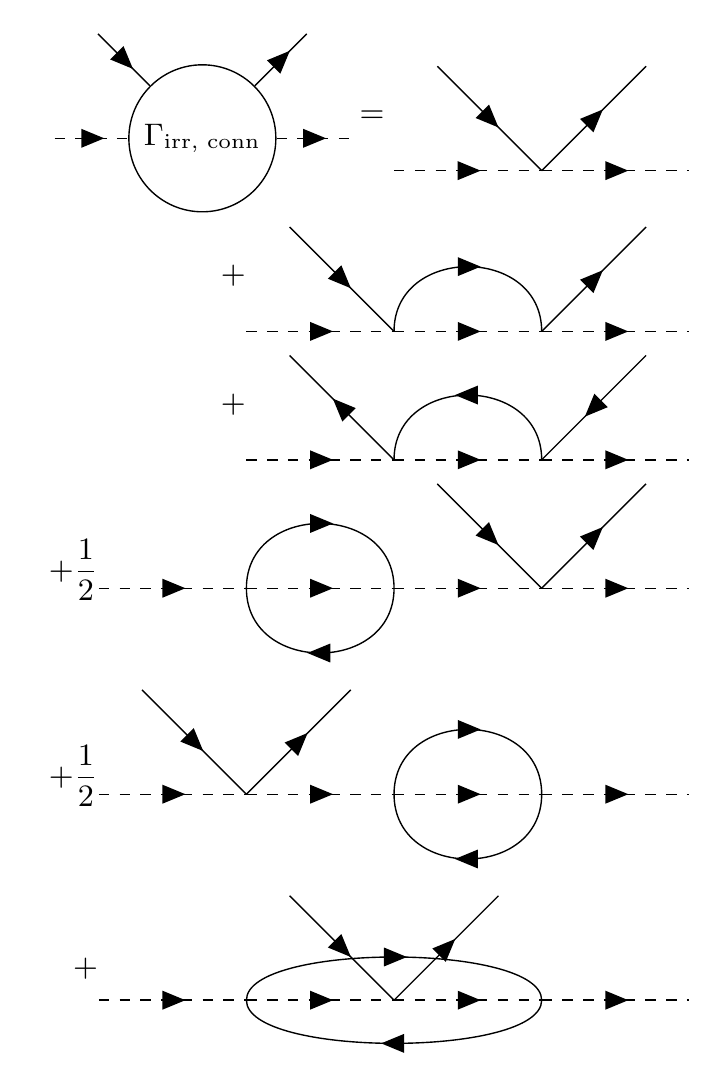}
\caption{Feynman diagrams contributing to the renormalization of interaction vertices at two loop order. We note that the two wave function renormalization diagrams are distinct for the case of anisotropic $\SU(N)$ Kondo Hamiltonians.} \label{fig:feynman}
\end{figure}

\section{ \label{app:betas} Solution of $\beta$-Functions}
The $\beta$-functions governing the flow of the Kondo couplings are enumerated in Section \ref{sec:rg}. Here, we present the fixed points $\bm{g}_*$ of the flow, which correspond to $\bm{\beta}(\bm{g}_*) = 0$.

\subsection{\label{app:b1g_betas} $A_{1g}\otimes B_{1g}$ Model}
We list here the renormalization group flow equations for the coupling constants in Eqs. \eqref{eq:HM1}-\eqref{eq:HO2}. 

\noindent The solutions are then given by $(J_{Q1}, J_{Q2}, J_{O1}) = \{(1,1,-1), (1,-1,1), (-1,1,1), (-1,-1,-1)\}$, and $J_M$ is free. As mentioned in the main text, each of these fixed points yields the same (up to unitary transformation) fixed point. In the main text, we have chosen the last of these four fixed points, and $J_M = 1$. 

\subsection{\label{app:a2g_betas} Combined model}
The $\beta$-functions for the model described in Sec. \ref{sec:a2g_model} are given by Eqs. \eqref{eq:a2g_beta1} - \eqref{eq:a2g_beta7} have solutions given by the following:

\begin{align}
J_{M} =& \eta,\\
J_{D1} =& \frac{2}{29} + \frac{5}{29}\zeta, \\
J_{D2} =& \frac{5}{29}\nu, \\
J_{Q1} =& -1, \\
J_{Q3} =& -1, \\
J_{Q2} =& -1, \\
J_{O3} =& -\frac{5}{29} + \frac{2}{29}\zeta, \\
J_{O1} =& - 1, \\
J_{O2} =& - 1, \\
J_{O4} =& \frac{2}{29} \nu .
\end{align}

\noindent As mentioned in the main text, $\eta,\zeta,\nu$ are all free parameters and can take any value, while still being a fixed point of the $\beta$-functions. The $\beta$-functions also have other solutions (corresponding to changing signs of some of the parameters) but all are equivalent up to unitary transformation.

\section{Strong Coupling Analysis}
In this section, we present the ground states resulting from the strong coupling analyses done on the two models of interest in Secs. \ref{sec:b1g_b2g_fixed_points} and \ref{sec:combined_model_analysis}. We recall that a perturbative intermediate fixed point is only stable if the corresponding strong coupling limit is unstable. An unstable strong coupling limit is signified by a degenerate ground state. The strong coupling ground state also needs to remain degenerate when perturbatively coupled to the conduction electrons \cite{Nozieres1980a}.

\subsection{\label{app:SC_b1g} Strong Coupling Analysis for $A_{1g}\otimes B_{1g}$ Model}
For the 2-channel $\SU(4)$ moment model, we calculate the strong coupling ground states in the $\chi$ conduction electron basis. There are between 0 and 4 conduction electrons able to occupy the impurity site, and, diagonalizing the fixed point Hamiltonian in all particle sectors, the doubly occupied sector has the lowest energy, with a 4-fold degenerate ground state. Here, the $\ket{\chi}$ states are defined by Eqs. \eqref{eq:SOC_basis_1} - \eqref{eq:SOC_basis_4} and the $\ket{\varphi}$ states are defined by Eqs. \eqref{eq:pseudo_basis_1} - \eqref{eq:pseudo_basis_4}. The ground states are given by:

\begin{align}
\ket{\text{GS}_1} =& -\frac{1}{\sqrt{6}}\ket{\chi_1\chi_3}\ket{\varphi_1} - \frac{1}{\sqrt{2}}\ket{\chi_2\chi_4}\ket{\varphi_1} + \sqrt{\frac{2}{3}}\ket{\chi_1\chi_4}\ket{\varphi_2}, \\
\ket{\text{GS}_2} =& -\sqrt{\frac{2}{3}}\ket{\chi_2\chi_3}\ket{\varphi_1} + \frac{1}{\sqrt{6}}\ket{\chi_1\chi_3}\ket{\varphi_2} + \frac{1}{\sqrt{6}}\ket{\chi_2\chi_4}\ket{\varphi_2}, \\
\ket{\text{GS}_3} =& -\frac{1}{\sqrt{6}}\ket{\chi_1\chi_3}\ket{\varphi_3} - \frac{1}{\sqrt{6}}\ket{\chi_2\chi_4}\ket{\varphi_3} + \sqrt{\frac{2}{3}}\ket{\chi_2\chi_3}\ket{\varphi_4}, \\
\ket{\text{GS}_4} =& \sqrt{\frac{2}{3}}\ket{\chi_1\chi_4}\ket{\varphi_3} - \frac{1}{\sqrt{6}}\ket{\chi_1\chi_3}\ket{\varphi_4} - \frac{1}{\sqrt{6}}\ket{\chi_2\chi_4}\ket{\varphi_4}.
\end{align}

\subsection{\label{app:SC_a2g} Strong Coupling Analysis for Combined Model}

For the combined model of moments in the $A_{1g}$, $A_{2g}$, $B_{1g}$, $B_{2g}$ models, the fixed point Hamiltonian was a truncated $\SU(4)$ Kondo model with 6 generators of $\SU(4)$. As in the previous section, the impurity site can host 0 to 4 conduction electrons. Diagonalizing the fixed point Hamiltonian in all particle sectors, the single particle sector has a 2-fold ground state, the two particle sector has a 4-fold ground state, and the three particle sector has a 2-fold ground state. The ground state energy in each of the different sectors is the same, so the total degeneracy of the ground state across all particle sectors is 8. In the single particle sector, the ground states are given by

\begin{align}
\ket{\text{GS}_1} = -\frac{1}{\sqrt{2}}\ket{\chi_1}\ket{\varphi_2} + \frac{1}{\sqrt{2}}\ket{\chi_2}\ket{\varphi_1}, \\
\ket{\text{GS}_2} = -\frac{1}{\sqrt{2}}\ket{\chi_3}\ket{\varphi_4} + \frac{1}{\sqrt{2}}\ket{\chi_4}\ket{\varphi_3}.
\end{align}

\noindent In the two particle sector, the ground states are given by 

\begin{align}
\ket{\text{GS}_1} =& \frac{1}{\sqrt{2}}\ket{\chi_1\chi_4}\ket{\varphi_3} - \frac{1}{\sqrt{2}}\ket{\chi_1\chi_3}\ket{\varphi_4}, \\
\ket{\text{GS}_2} =& -\frac{1}{\sqrt{2}}\ket{\chi_1\chi_4}\ket{\varphi_2} + \frac{1}{\sqrt{2}}\ket{\chi_2\chi_4}\ket{\varphi_1}, \\
\ket{\text{GS}_3} =& -\frac{1}{\sqrt{2}}\ket{\chi_1\chi_3}\ket{\varphi_2} + \frac{1}{\sqrt{2}}\ket{\chi_2\chi_3}\ket{\varphi_1}, \\
\ket{\text{GS}_4} =& \frac{1}{\sqrt{2}}\ket{\chi_2\chi_4}\ket{\varphi_3} - \frac{1}{\sqrt{2}}\ket{\chi_2\chi_3}\ket{\varphi_4} .
\end{align}

\noindent In the three particle sector, the ground states are given by

\begin{align}
\ket{\text{GS}_1} =& -\frac{1}{\sqrt{2}}\ket{\chi_1\chi_3\chi_4}\ket{\varphi_2} +\frac{1}{\sqrt{2}}\ket{\chi_2\chi_3\chi_4}\ket{\varphi_1}, \\
\ket{\text{GS}_2} =& \frac{1}{\sqrt{2}}\ket{\chi_1\chi_2\chi_4}\ket{\varphi_3} - \frac{1}{\sqrt{2}}\ket{\chi_1\chi_2\chi_3}\ket{\varphi_4}.
\end{align}

\section{\label{sec:effective_moments} Effective Moment Operators}
The pseudospin operators in Section \ref{sec:combined_model_analysis} are explicitly shown here in the basis given by Eqs. \eqref{eq:pseudo_basis_1} -\eqref{eq:pseudo_basis_4}. The $\sigma^i$ represent canonically normalized $\algsu(2)$ Pauli matrices, and the 0's are $2\times 2$ zero matrices.

\begin{align}
S^8+S^{14} =& 2\begin{pmatrix} \sigma^x & 0 \\ 0 & 0 \end{pmatrix} = 2\lambda^1 \label{eq:eff_moment_1} \\
S^4+S^{10} =& 2\begin{pmatrix} \sigma^y & 0 \\ 0 & 0 \end{pmatrix} = 2\lambda^2 \label{eq:eff_moment_2}\\
S^0-2S^2 - S^6-S^{12} =& 2\begin{pmatrix} \sigma^z & 0 \\ 0 & 0 \end{pmatrix} = 2\lambda^3 \label{eq:eff_moment_3} \\
S^8-S^{14} =& 2\begin{pmatrix} 0 & 0 \\ 0 & \sigma^x \end{pmatrix} = 2\lambda^{13} \label{eq:eff_moment_4}\\
S^4-S^{10} =& 2\begin{pmatrix} 0 & 0 \\ 0 & \sigma^y \end{pmatrix} = 2\lambda^{14} \label{eq:eff_moment_5}\\
-S^0 - 2S^2 + S^6 - S^{12} =& 2\begin{pmatrix} 0 & 0 \\ 0 & \sigma^z \end{pmatrix} = 2\tilde{\lambda} \label{eq:eff_moment_6}
\end{align}

\end{appendix}


%

\end{document}